\documentclass{kluwer}
\usepackage{graphicx}  

\begin{document}
\begin{article}
\begin{opening}
\title{The Connection between Spheroidal Galaxies and QSOs}

\author{G.L. \surname{Granato}\email{granato@pd.astro.it}}
\author{G. \surname{DeZotti}}
\institute{Osservatorio Astronomico di Padova}
\author{L. \surname{Silva}}
\institute{Osservatorio Astronomico di Trieste}
\author{L. \surname{Danese}}
\author{M. \surname{Magliocchetti}}
\institute{SISSA - Trieste}



\runningtitle{QSOs and Spheroidal Galaxies}
\runningauthor{Granato e.a.}


\begin{abstract} 
In view of the extensive evidence of tight inter-relationships
between spheroidal galaxies (and galactic bulges) with massive
black holes hosted at their centers, a consistent model must deal
jointly with the evolution of the two components. We describe one
such model, which successfully accounts for the local luminosity
function of spheroidal galaxies, for their photometric and
chemical properties, for deep galaxy counts in different
wavebands, including those in the (sub)-mm region which proved to
be critical for current semi-analytic models stemming from the
standard hierarchical clustering picture, for clustering
properties of SCUBA galaxies, of EROs, and of LBGs, as well as for
the local mass function of massive black holes and for quasar
evolution. Predictions that can be tested by surveys carried out
by SIRTF are presented.
\end{abstract}


\abbreviations{\abbrev{KAP}{Kluwer Academic Publishers};
   \abbrev{compuscript}{Electronically submitted article}}

\nomenclature{\nomen{KAP}{Kluwer Academic Publishers};
   \nomen{compuscript}{Electronically submitted article}}

\classification{JEL codes}{D24, L60, 047}
\end{opening}

\section{Introduction}
The hierarchical clustering model with a scale invariant spectrum
of density perturbations in a Cold Dark Matter (CDM) dominated
universe has proven to be remarkably successful in matching the
observed large-scale structure as well as a broad variety of
properties of galaxies of the different morphological types 
(Granato et al. 2000 and references therein). Serious
shortcomings of this scenario have also become evident in recent
years. 

At the other extreme of the galaxy mass function with respect to
so-called ``small-scale crisis'', another strong discrepancy
with model predictions arises, that we might call ``the massive galaxy
crisis''. Even the best semi-analytic models hinging upon the standard
picture for structure formation in the framework of the hierarchical
clustering paradigm, are stubbornly unable to account for the (sub)-mm
(SCUBA, see Fig.~1, and MAMBO) counts of galaxies, most of which are
probably massive objects undergoing a very intense star-burst (with star
formation rates $\sim 1000\,\hbox{M}_\odot\,\hbox{yr}^{-1}$) at $z>2$.
Recent optical data confirm that most massive
ellipticals were already in place and (almost) passively evolving up to
$z\simeq 1$--1.5. These data are more consistent with the traditional
``monolithic'' approach whereby giant ellipticals formed most of their
stars in a single gigantic starburst at substantial redshifts, an
underwent essentially passive evolution thereafter.

In the canonical hierarchical clustering paradigm the smallest objects
collapse first and most star formation occurs, at relatively low rates,
within relatively small proto-galaxies, that later merged to form larger
galaxies. Thus the expected number of galaxies with very intense star
formation is far less than detected in SCUBA and MAMBO surveys and the
surface density of massive evolved ellipticals at $z\gsim 1$ is also
smaller that observed. The ``monolithic'' approach, however, is
inadequate to the extent that it cannot be fitted in a consistent
scenario for structure formation from primordial density fluctuations.

\section{Relationships between quasar and galaxy evolution}
The above difficulties, affecting even the best current recipes,
may indicate that new ingredients need to be taken into account. A
key new ingredient may be the mutual feedback between formation
and evolution of spheroidal galaxies and of active nuclei residing
at their centers. In this framework, Granato et al. (2001)
elaborated the following scheme.

Feed-back effects, from supernova explosions and from active nuclei
delay the collapse of baryons in smaller clumps while large ellipticals
form their stars as soon as their potential wells are in place; {\it the
canonical hierarchical CDM scheme -- small clumps collapse first -- is
therefore reversed for baryons}. Large spheroidal galaxies therefore
undergo a phase of high (sub)-mm luminosity.

At the same time, the central black-hole (BH) grows by accretion and the
quasar luminosity increases; when it reaches a high enough value, its
action stops the star formation and eventually expels the residual gas.
The same mechanism distributes in the inter-galactic medium a
substantial fraction of metals. The duration of the star-burst, imposed
by the onset of quasar activity, increases with decreasing mass from
$\sim 0.5$ to $\sim 2\,$Gyr.

This implies that the star-formation activity of the most massive
galaxies quickly declines for $z\lsim 3$, i.e. that the redshift
distribution of SCUBA/MAMBO galaxies should peak at $z\gsim 3$, as
quasars reach their maximum luminosity (at $z\simeq 2.5$). This
explains why very luminous quasars are more easily detected at (sub)-mm
wavelengths for $z \gsim 2.5$. 

A ``quasar phase'' follows, lasting $10^7$--$10^8\,$yrs, and a long
phase of passive evolution of galaxies ensues, with their colors
becoming rapidly very red [Extremely Red Object (ERO) phase].
Intermediate- and low-mass spheroids have lower Star Formation Rates
(SFRs) and less extreme optical depths. They show up as Lyman-Break
Galaxies (LBGs).

Therefore, in this scenario, large ellipticals evolve essentially as in
the ''monolithic'' scenario, yet in the framework of the standard
hierarchical clustering picture. Many aspects and implications of this
compound scheme have been addressed by our group in a series of papers
(Granato et al.\ 2001, Magliocchetti et al.\ 2001, Perrotta et al.\
2001, Romano et al.\ 2001). Here we only summarize how the scenario
compare with sub-mm counts.

\begin{figure} 
\begin{center}
\includegraphics[width=1.0\textwidth]{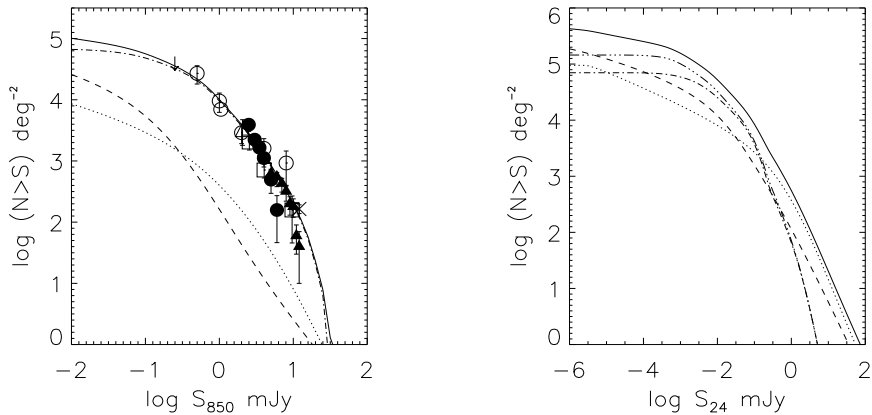}
\end{center}
\caption[]
{Left hand panel: integral source counts at $850\,\mu$m
 predicted by the model
by Granato et al.\ 2001 compared with observations. The dotted,
dashed, dot-dashed and solid lines show the contributions of starburst,
spiral, forming elliptical galaxies, and total respectively. 
Right hand panel: integral source counts 
predicted by the same model. The dotted, dashed and
dot-dashed lines show the contributions of starburst, spiral, and
forming elliptical galaxies, respectively, while the
three-dots/dashed line shows the total counts of ellipticals,
including also those where the star-formation has ended.
}
\label{figura}
\end{figure}

\subsection{Counts at (sub)-mm wavelengths}
The (sub)-mm counts are expected to be very steep because of the
combined effect of the strong cosmological evolution of dust emission in
spheroidal galaxies and of the strongly negative K-correction (the dust
emission spectrum steeply rises with increasing frequency). The model by
Granato et al. (2001) has extreme properties in this respect: above
several mJy its $850\,\mu$m counts reflect the high-mass exponential
decline of the mass function of dark halos. In this model, SCUBA/MAMBO
galaxies correspond to the phase when massive spheroids formed most of
their stars at $z\gsim 2.5$; such objects essentially disappear at lower
redshifts. On the contrary, the counts predicted by alternative models
(which are essentially phenomenological) 
while steep, still have a power law
shape, and the redshift distribution has an extensive low-$z$ tail.
As illustrated by Fig.~1, the recent relatively large area surveys
are indeed suggestive of an
exponential decline of the $850\,\mu$m counts above several mJy.
Further evidence in this direction comes from MAMBO surveys at
$1.2\,$mm.

\subsection{Predictions for SIRTF surveys}
SIRTF surveys have the potential of providing further tests of the
model. In particular the $24\,\mu$m survey to be carried out as a
part of the GOODS (http://www.stsci.edu/science/goods) Legacy
Science project should reach a flux limit of $100\,\mu$Jy.
According to the model, about 50\% of detected galaxies should be
spheroidal galaxies forming their stars at $z \gsim 2$. About
400--600 such objects are expected over an area of 0.1 square
degree (see Fig.~2). Their redshift distribution is predicted to
peak at $z$ slightly above 2, with a significant tail extending up
to $z\simeq 3$.

\acknowledgements We benefited from many helpful exchanges with C.
Baccigalupi, F. Matteucci, F. Perrotta, D. Romano. Work supported in
part by ASI and MIUR. G.L.G. thanks SADG for partial financial support.

\end{article}

\begin{thebibliography}{}


\bibitem{Magliocchetti2001} M. Magliocchetti, L. Moscardini,
P. Panuzzo, G.L. Granato, G. De
Zotti, L. Danese: MNRAS, \textbf{325}, 1553 (2001)

\bibitem{Granato2000} G.L. Granato, C.G. Lacey,
L. Silva, A. Bressan, C.M. Baugh, S. Cole,
C.S. Frenk, C. S.: ApJ \textbf{542}, 710 (2000)

\bibitem{Granato2001} G.L. Granato, L. Silva,
P. Monaco, P. Panuzzo, G. De
Zotti, L. Danese, L. 2001: MNRAS \textbf{324}, 757 (2001)

\bibitem{Perrotta2001b} F. Perrotta, M. Magliocchetti et al.: MNRAS,
submitted, astro-ph/0111239 (2001)

\bibitem{Romano2001} D. Romano, L. Silva, F. Matteucci, L. Danese:
preprint (2001)

\end{thebibliography}
\end{document}